# Analysis of $Nb_3Sn$ surface layers for superconducting RF cavity applications

Chaoyue Becker,[1,2,3] Sam Posen,[4] Nickolas Groll,[1,2] Russell Cook,[5] Christian M. Schlepütz,[6] Daniel Leslie Hall,[4] Matthias Liepe,[4,7] Michael Pellin,[1] John Zasadzinski,[3] and Thomas Proslier[1,2,a)]

[1)] *Materials Science Division, Argonne National Laboratory, Argonne, Illinois 60439, USA*
[2)] *High Energy Physics Division, Argonne National Laboratory, Argonne, Illinois 60439, USA*
[3)] *Department of Physics, Illinois Institute of Technology, Chicago, Illinois 60616,USA*
[4)] *Cornell Laboratory for Accelerator-Based Sciences and Education, Ithaca, NY 14853, USA*
[5)] *Nanoscience and Technology Division, Argonne National Laboratory, Argonne, Illinois 60439, USA*
[6)] *X-ray Science Division, Argonne National Laboratory, Argonne, Illinois 60439, USA*
[7)] *Department of Physics, Cornell University, Ithaca, NY 14853, USA*

We present an analysis of $Nb_3Sn$ surface layers grown on a bulk Niobium (Nb) coupon prepared at the same time and by the same vapor diffusion process used to make $Nb_3Sn$ coatings on 1.3 GHz Nb cavities. Tunneling spectroscopy reveals a well-developed, homogeneous superconducting density of states at the surface with a gap value distribution centered around 2.7 ±0.4 meV and superconducting critical temperatures ($T_c$) up to 16.3K. Scanning transmission electron microscopy (STEM) performed on cross sections of the sample's surface region shows a ~ 2 microns thick $Nb_3Sn$ surface layer. The elemental composition map exhibits a Nb:Sn ratio of 3:1 and reveals the presence of buried sub-stoichiometric regions that have a ratio of 5:1. Synchrotron x-ray diffraction experiments indicate a polycrystalline $Nb_3Sn$ film and confirm the presence of Nb rich regions that occupy about a third of the coating volume. These low $T_c$ regions could play an important role in the dissipation mechanisms occurring during RF tests of $Nb_3Sn$ -coated Nb cavities and open the way for further improving a very promising alternative to pure Nb cavities for particle accelerators.

Discovered in 1954[1], the A-15 compound $Nb_3Sn$ is a Type II ($\kappa$~20) strong coupling s-wave superconductor[2,3] with a maximum $T_c$ of 18 K[4] and superconducting order parameter $\Delta$ of 3.4 meV[5]. Due to its relatively high $T_c$ and ability to carry high current densities, $Nb_3Sn$ is an ideal candidate for replacing NbTi for superconducting wire applications and Nb for superconducting radio frequency (SRF) resonators operating from a few hundred MHz up to several GHz. Early work into developing $Nb_3Sn$ for SRF applications started in the 1970s[6–9]. In particular, researchers from Wuppertal University optimized a coating recipe[8] based on the diffusion of Sn vapor into elemental Nb at temperatures between 1000°C to 1200°C. This approach has the unique advantage of being scalable to applications for which a coating process without a direct line of sight is required. State-of-the-art RF performance tests then[10] showed an extremely high quality factor ~ $10^{11}$ at 2K and ~ $10^{10}$ at 4.2K (about 20 times higher than pure Nb) with a strong decrease of the quality factor ($Q$-slope) above an accelerating field of 5 MV/m. The origin of this $Q$-slope remains unclear, however it was postulated that the onset of the Q decrease at 5 MV/m (peak surface magnetic field of 22 mT) was due to early vortex penetration above the $Nb_3Sn$ first critical field $B_{C1}$, and therefore was an intrinsic material limitation. A regain of interest was stimulated by recent RF tests done at Cornell University[11] that reproducibly exhibit a similar $Q$ factor ~ $2 \times 10^{10}$ at 4.2K (and $3 \times 10^{10}$ at 2K), but a very moderate $Q$-slope up to a quenching field of 12-17 MV/m, corresponding to a peak surface magnetic field of 50-70 mT, which is significantly higher than the $B_{C1}$ of 25±7 mT measured on this cavity[11]. Another striking and reproducible feature is the very moderate increase of the Q factor at low field between 4.2K and 2K for previous and recent Cornell RF tests. This saturation of the Q factor in temperature is the signature of a residual, non-BCS dissipation mechanism that dominates below 4.2K.

In order to investigate the microscopic origins of potential defect-induced performance limitations, we used various surface characterization techniques on the same $Nb_3Sn$ sample provided by Cornell University. This sample was subject to the same Wuppertal deposition process used to coat Nb cavities and was prepared at the same time as a $Nb_3Sn$ cavity that shows the characteristic RF performance features mentioned above. For more details on the process method see Ref.11.

The scanning electron microscopy (SEM) secondary-electron image shown in Fig. 1(a) reveals the presence of well-defined grains with an area distribution centered around 0.6 $\mu m^2$, which corresponds to an average grain size of 0.9 $\mu$m. The room-temperature x-ray diffraction measurements represented in Fig. 1(b) and (c) were collected at the Advance Photon Source (APS, beamline 33-BM) at an energy of 15.5±0.0015 keV. The symmetric $\theta$-$2\theta$ scan unveils the presence of both $Nb_3Sn$ and Nb diffraction peaks, establishing that the entire $Nb_3Sn$ film volume is probed. The $Nb_3Sn$ peak intensities closely match the powder diffraction reference and the recorded diffraction images on the Pilatus 100K pixel detector (not shown) show rings of constant intensity, both of which are strong evidence of a polycrystalline film. A closer look at the $Nb_3Sn$ diffraction peaks reveals that each peak is asymmetric with a shoulder at higher $2\theta$ values that can

[a)] Electronic mail: prolier@anl.gov

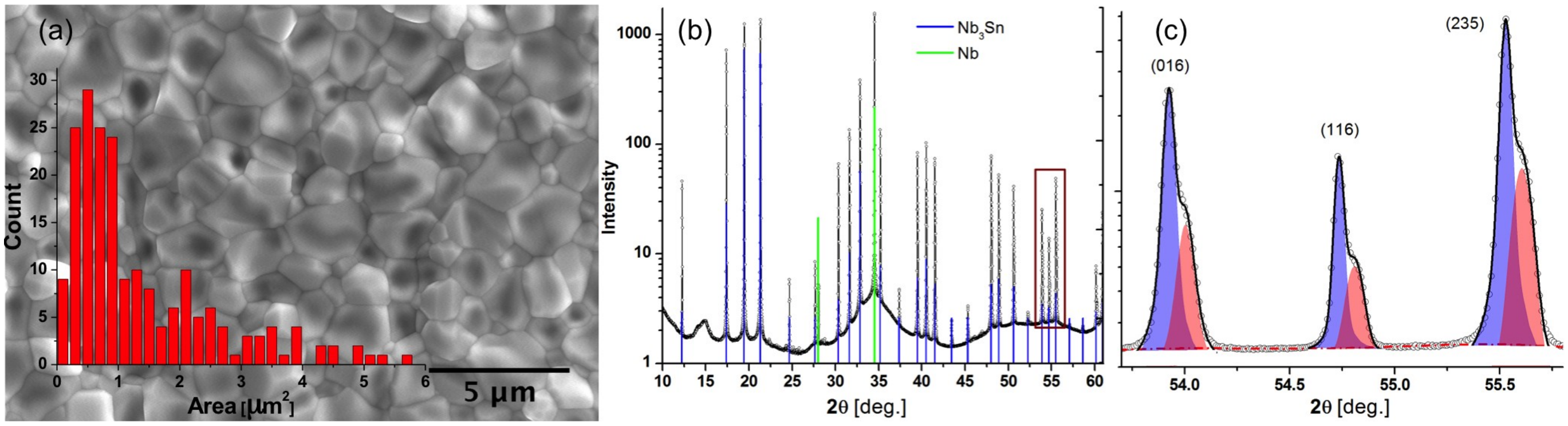

FIG. 1. (Color online) Structural characterization of a $Nb_3Sn$ film grown by the Wuppertal method on a cavity-grade Nb sample. A typical SEM image is shown in (a) - the insert displays the averaged statistical grain area distribution extracted from more SEM pictures (not shown). (b) Diffraction pattern of a $Nb_3Sn$ film on top of bulk Nb (open dots) and the corresponding powder diffraction reference patterns for $Nb_3Sn$ (ICSD 105230, blue) and Nb (ICSD 76554, green). The box drawn in (b) indicates the zoom area represented in (c) that reveals the peak splitting observed for all $Nb_3Sn$ diffraction peaks. The fits using Voigt functions are shown in blue and red. $2\theta$ represents the angle between the incident X-rays beam and the detector.

be fitted with 2 Voigt functions as shown in Fig. 1(c). It is noteworthy to mention that the Nb peaks do not show any asymmetric shoulders, thus ruling out a potential instrument artifact. The Nb peaks positions indicate an unperturbed cubic phase with a lattice parameter of $a$ = 3.303 Å. The peaks shown in blue match a stoichiometric cubic $Nb_{1-\delta}Sn_{\delta}$ phase with $\delta = 0.25 \pm 0.01$ and a lattice parameter of $a$ = 5.29 Å, whereas the red peaks correspond to a reduced lattice parameter of $a$ = 5.282 Å which, according to the work of Devantay *et al.*[12], indicates a Nb rich A-15 phase $Nb_{1-\delta}Sn_{\delta}$ with $\delta = 0.18 \pm 0.01$. Following the approach of Williamson and Hall[13], we extracted an average crystallite size of $d = 290 \pm 30$ nm and a strain of $E = 0.027 \pm 0.002\%$ for the "blue" phase and $E = 0.1 \pm 0.01\%$ for the "red" phase. The peak area ratio of the two stoichiometries is consistently found to be 0.36±0.04 for all peaks, suggesting that the low Sn A-15 phase takes up 25±3% of the film volume.

In order to further examine the spatial distribution of these two stoichiometries, we performed an elemental composition study by STEM and XEDS (X-ray energy dispersive spectroscopy) on two cross sections prepared by focused ion beam (Zeiss 1540XB FIB) and argon ion milling of the $Nb_3Sn$ film. The quantitative analysis of both cross sections gave very similar results and one of them is represented in Fig.2. The TEM model used is a Tecnai F20ST at 200 kV in STEM mode and the XEDS map frame times were 25 s to eliminate scanning fly-back problems. Each map consists of 200 x 256 points with a point spacing of ~ 10 nm, which is larger than the 2 nm beam size to avoid overlapping data from neighboring pixels. The XEDS data at each point is an average over the cross section thickness estimated to be ≤ 50 nm in the sampling area. The field of view shows large and well-defined grains of several microns, consistent with the SEM image, most of them extending all the way through the film thickness. The regions labeled as 2 are a few hundred nm in size (between 150 to 550 nm) with a composition of $Nb_{1-\delta}Sn_{\delta}$ with $\delta = 0.17 \pm 0.002$, surrounded by areas with $\delta = 0.25 \pm 0.0025$ labeled as 1 in Fig.2. On both elemental maps measured, regions 2 are mostly localized near the Nb/$Nb_3Sn$ interface. Isolated poaches of region 2 are also present closer to the surface at a depth between 0.8 to 0.35 $\mu$m. The compositions of the two types of regions are in very good agreement with the diffraction data. Such regions show very strong composition gradients on relatively short length scales ~ 50 nm, which can explain the higher strain of the Nb rich A-15 phase measured by XRD. The area ratio between region 1 and 2 is 0.35 ±0.02, also in agreement with the diffraction data. The average grain size of the stoichiometric $Nb_3Sn$ phase extracted from XRD (290 nm), however, is almost an order of magnitude smaller than that measured with the SEM (~ 1$\mu$m). This apparent discrepancy can be explained by understanding that the SEM secondary electron image derives from a non-diffraction process and only shows the surface of grains that are separated by well-defined grain boundaries, whereas XRD is sensitive to the presence of sub-grain structures such as local mosaicity within the volume of individual grains. Upon looking more closely at the STEM image (Fig. 2(a)), we can actually distinguish smaller structures with characteristic sizes of around 200 nm within the large grains.

The presence of only two compositions for the A-15 phase (25% and 17%) observed in XRD and STEM-XEDS maps is consistent with the Nb-Sn intermetallic phase diagram[12,22]. The phase diagram indicates that at 1000°C if the Sn content drops below 17% (which is bound to happen during a diffusion process) then one enters a mixed phase region consisting of pure Nb and the A-15 phase at 17%. Such lower average Sn content areas would lead to a two-phase segregation and a pile-up of regions with 17% Sn in the A-15 phase.

The cavity $Q$ factor is extremely sensitive to the surface superconducting gap $\Delta$ and $T_c$ . We measured the sur-face superconducting properties by means of point contact tunneling (PCT) spectroscopy. In the tunnel regime, the differential conductance is directly proportional to the superconducting density of states, $N_S(E)$, through the relation:

$$\frac{dI_{NS}}{dV} = \int_{-\infty}^{+\infty} N_s(E,\Gamma,\Delta)\left[-\frac{\partial f(E+eV)}{\partial(eV)}\right] dE, (1)$$

where $I_{NS}$ is the current flowing between the normal metal $N$ (tip) and the superconductor $S$ (sample) under a difference of potential $V$, $f(E)$ is the Fermi function, $\Gamma$ is the phenomenological quasi-particle lifetime broadening parameter[14] and $\Delta$ is the superconducting gap. Tunneling spectroscopy probes the superconducting properties within a depth of few coherent length, $\xi$, of the surface ($\xi \sim 3-6$ nm[15,16]). In our setup, the junctions are made by approaching the sample surface with a PtIr tip, creating a SIN (superconductor-insulator-normal) junction where an oxide layer serves as the insulator. The sample is mounted on a X-Y-Z piezo stage that control both the tip-sample separation and the lateral positioning, thus enabling large scale mapping of the surface density of states[17].

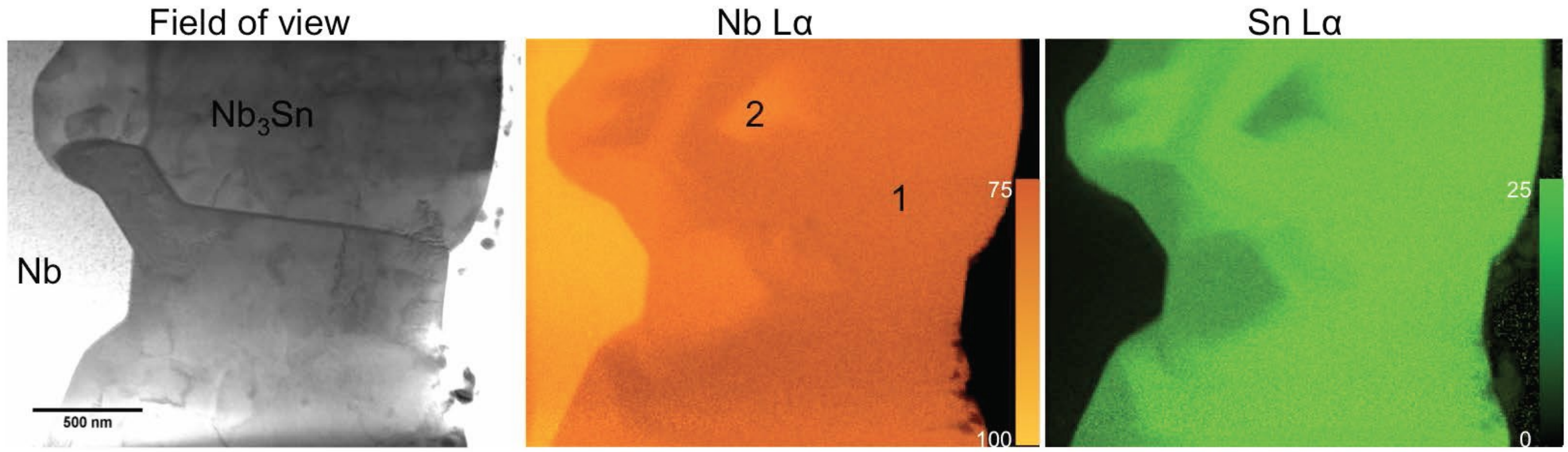

FIG. 2. (Color online) Cross sectional images of the $Nb_3Sn$ coating on a bulk Nb sample. The elemental maps for the Nb and Sn L$\alpha$ XEDS peaks were obtained by normalizing the corresponding peak areas to the background for the XEDS spectrum in each point of the map. The experimental k-factors for the Nb and Sn L$\alpha$ emission lines were derived assuming a composition of 25 atomic % for the higher-Sn (consistent with XRD) region from the $Nb_3Sn$ phase diagram. Those k-factors were then applied to a subsequent analysis of the XEDS data from the low-Sn regions. Regardless of the analysis model used, there is always 8% difference between the low and high Sn regions. The scale bar units are in atomic %.

Typical normalized conductance spectra as shown in Fig. 3 reveal a well-developed superconducting gap around the Fermi level (V=0 mV). The distribution and maps of $\Delta$ and $\Gamma/\Delta$ displayed in Fig. 4 were extracted from the fits of the tunneling conductance spectra measured at 100 junctions distributed over an area of 800 x 400 $\mu m^2$ by using equation (1). The ratio $\Gamma/\Delta$ reflects the value of the normalized conductance at the Fermi level and, as such, should be a relevant parameter to correlate to the dissipation occurring in transport. A weak spot localized around X= 100 $\mu$m and Y=-300 $\mu$m is clearly visible in the maps and seems to be correlated with a carbon rich region measured by XEDS (not shown) that appears as a black spot in the SEM image in Fig.

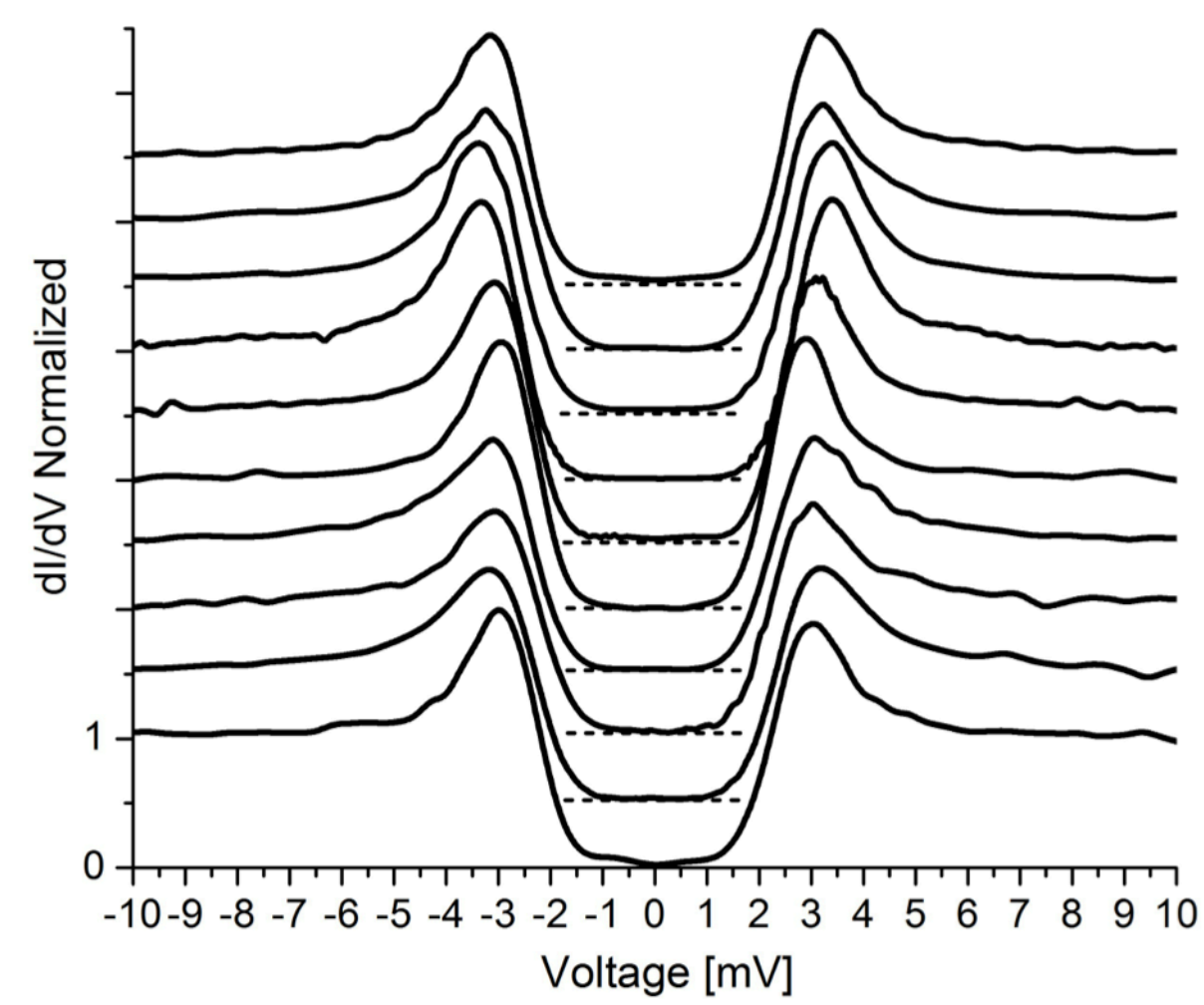

FIG. 3. Characteristic normalized tunneling conductance spectra measured on the $Nb_3Sn$ surface. The spectra have been offset vertically by 0.5 for clarity. The horizontal dashed lines correspond to the zero conductance value for each of the curves.

4(a). It is unclear at this point whether this speck is due to post-contamination or was already present at an earlier stage of the Nb processing. There is considerable evidence[18], however, of carbon-rich regions developing on the surface of processed Nb that are commonly found to be $10-20\mu$m across, $\sim 1\mu$m deep and impervious to high temperature treatments. The $Nb_3Sn$ growth process may not completely cover or remove these regions with a significant carbon excess. The $\Delta$ and $\Gamma/\Delta$ values obtained in this region were not included in the statistics represented in Fig. 4(d) and the discussion is focused on the $Nb_3Sn$ .

Overall, both the gap $\Delta$ and the ratio $\Gamma/\Delta$ appear to be uniform on the sampled area with an average of $\Delta = 2.7 \pm 0.4$ meV (nearly twice the Nb gap) in agreement with a $\delta = 0.25$[5]. The values of $\Gamma/\Delta = 3.6 \pm 5\%$ are generally much smaller than the typical values found on cavity grade Nb samples (~ 10-20%)[19–21] and therefore are indicative of low quasiparticle dissipation. The large gap values and the low quasiparticle dissipation are in agreement with the high quality factors measured at 4.2K. The temperature dependence of two junctions (not shown) with superconducting gap values of $\Delta(0) =$ 2.89 meV and 2.54 meV around the peak distribution (Fig. 4(d)) reveal a classical BCS temperature dependence with a $T_c$ of 16.2K, $2\Delta(0)/k_BT_c$ of 4.16, and a $T_c$ of 13.75K with a $2\Delta(0)/k_BT_c$ of 4.28, respectively. The presence of areas with $\Delta \leq 2.5$meV and $T_c \leq 13.75$K located around Y = 100 $\mu$m indicate some mechanism reducing the superconducting properties below those of the phase with $\delta = 0.25$. The regions with $\delta = 0.17 \pm 0.01$ are known to have poor superconducting properties with a reported $T_c$ ~ 6K and superconducting gap $\Delta$ ~ 1meV[22]. It is possible that the variation of $T_c$ and $\Delta$ can be due to a proximity effect induced by these low $T_c$ regions into the surrounding high $T_c$ ones. In the proximity effect model[23], the superconducting gap parameter varies on a length scale of $\xi$ in the high $T_c$ superconducting phase, implying that some of these Nb rich regions must be present within ~10 nm below the surface. Such low $T_c$ regions, while not probed directly by tunneling, are pulling down the measured gap value at the surface.

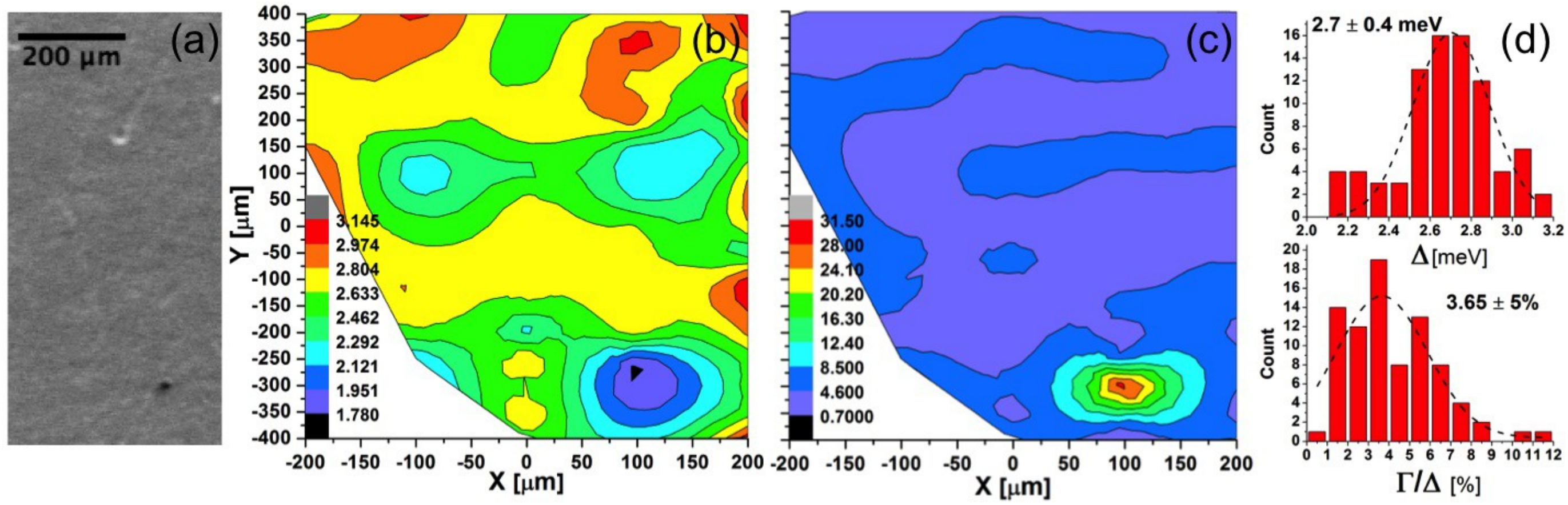

FIG. 4. (Color online) Map of the superconducting density of states measured over an area of 800x400 $\mu$m. An SEM image of the measured region is shown in (a). The color map represented in (b) is generated from the gap values (in meV) of the 100 junctions measured over the entire area of the sample. Gap values between adjacent dots are estimated by a linear interpolation. Regions of constant color have a constant gap value. The legend shows how color corresponds to the value of the gap. A similar map is shown for the parameter $\Gamma/\Delta$ in % in (c). The statistical distributions of these two parameters are represented in (d).

We will be now discussing the relevance of the surface analysis results for the RF performance of the Nb3Sn cavities. Under external RF magnetic field excitation, the measured quality factor $Q$ of the cavity (which is an average over the entire inner cavity surface) is $\propto e^{-\Delta/k_BT}$ and is dominated by dissipation mechanisms occurring within a few penetration depths, i.e. over a few hundred nm from the surface ($\lambda = 65 - 89$ nm[15,16] for the clean limit and 140 nm[24] for this cavity). The presence of Sn rich grain boundaries was postulated[11] as one possible reason for the linear increase of the surface resistance with magnetic field amplitude. We find no evidence of such sub-stoichiometry, but rather Nb rich grains with sizes of approximately 200 nm. These localized Nb rich regions with lower $T_c$ and $\Delta$ values would act as dissipative centers as well as providing easy entry points for magnetic flux effectively decreasing the overall Q factor. At this point the quantitative effect of such defects on the temperature and magnetic field dependence of $Q$ is unclear; such regions of reduced gap size however indicate that, despite the overall high Q of these cavities, we have not yet achieved maximum performance. Indeed, the data may be suggesting possible changes in processing. For example, it might be necessary to increase the Sn vapor concentration during diffusion to increase the average Sn concentration on the surface or to use a faster cool down rate at the end of the process that would freeze the Sn diffusion.

This work was funded by the U.S. Department of Energy, Office of Sciences, Office of High Energy Physics, early career award FWP 50335 and DOE award ER41628. Use of the Center for Nanoscale Materials and resources of the Advanced Photon Source was supported by the U. S. Department of Energy, Office of Science, Office of Basic Energy Sciences and Office of Science User Facility, under Contract No. DE-AC02-06CH11357.